\documentstyle[preprint,prl,aps,epsf]{revtex}
\begin{document}
\newcommand{\be}{\begin{eqnarray}}
\newcommand{\ee}{\end{eqnarray}}
\tightenlines
\vskip 2.5cm
\preprint{LBNL-56656}
\title{Shadowing and Absorption Effects on $J/\psi$ Production in $dA$
Collisions}

\author{R. Vogt}
 
\address{{
Nuclear Science Division, Lawrence Berkeley National Laboratory, 
Berkeley, CA 94720, USA\break 
and\break
Physics Department, University of California, Davis, CA 95616, USA}\break}
 
\vskip .25 in
\maketitle
\begin{abstract}
We study medium modifications of $J/\psi$ production in cold
nuclear media in deuterium-nucleus collisions.  We discuss several
parameterizations of the modifications of the parton densities in the nucleus,
known as shadowing, an initial-state effect.  We also include 
absorption of the produced $J/\psi$ by nucleons, a final-state effect.
Both spatially homogeneous and inhomogeneous shadowing and absorption are
considered.  We use the number of binary nucleon-nucleon collisions as a
centrality measure.  Results are presented for d+Au collisions at 
$\sqrt{S_{NN}} = 200$ GeV and for d+Pb collisions at $\sqrt{S_{NN}} = 6.2$ TeV.
To contrast the centrality dependence in $pA$ and d$A$ collisions, we also 
present $p$Pb results at $\sqrt{S_{NN}} = 8.8$ TeV.
\end{abstract}

\vskip 0.2in

The nuclear, or $A$, dependence of $J/\psi$ production is an important topic
of study for nuclear collisions.  It is essential that the $A$ dependence be
understood in cold nuclear matter to set a proper baseline for quarkonium
suppression in $AA$ collisions.  At fixed target energies, NA50 \cite{NA50}
has studied the $J/\psi$ $A$ dependence and attributed its behavior to $J/\psi$
break up by nucleons in the final state, referred to as nuclear absorption.
However, it is also known that the parton distributions are modified in
the nucleus relative to free protons.  This modification, referred to here as
shadowing, is increasingly important at higher energies.  This paper studies
the interplay of shadowing and absorption in deuteron-gold collisions at
the Relativistic Heavy Ion Collider (RHIC) and in deuteron-lead and proton-lead
collisions at the future Large Hadron Collider (LHC).  We consider both 
spatially homogeneous (minimum bias collisions) and inhomogeneous (fixed
impact parameter) results.  When possible, we discuss the results in the
context of data, in particular, the PHENIX data from RHIC.

The nuclear quark and antiquark distributions have been probed through
deep inelastic scattering (DIS) of leptons and neutrinos from nuclei.
These experiments showed that parton densities in free protons are
modified when bound in the nucleus \cite{Arn}.  At low parton momentum 
fractions, $x<0.02$, the ratio of the nuclear parton density relative 
to the nucleon is less than unity (the shadowing region).  In the intermediate
$x$ regime, $0.02< x < 0.1$, this ratio is larger than unity (antishadowing)
while at higher $x$ it drops below unity once more (the EMC region). 
The effect also depends on the scale of the interaction, the square of 
the momentum transfer, $Q^2$.  As the collisions energy increases, the values
of $x$ probed in the collision are decreased, making shadowing more important.
Since shadowing affects the parton distribution functions before the collision
that produces the $J/\psi$ occurs, it is an initial state effect.

Shadowing should
depend on the spatial position of the interacting parton
within the nucleus \cite{ekkv4}.  Although DIS
experiments are typically insensitive to this position dependence, 
some spatial inhomogeneity has been observed in $\nu N$
scattering \cite{E745}.  Thus the effect should be sensitive to the impact
parameter, $b$, at which the collision occurs so that the results depend on 
the collision centrality.  Central collisions with low impact parameter
should exhibit stronger shadowing effects than collisions in the nuclear 
periphery.  In Refs.~\cite{ekkv4,psidaprl}, we discussed the effects of
spatially inhomogeneous shadowing on $J/\psi$ production in $AA$ and d$A$
collisions respectively.  In this paper, we augment the d$A$ predictions by
including final-state absorption of the $J/\psi$ by nucleons.

At lower fixed-target energies, such as those at the CERN SPS, the $x$ values
are rather high, $x \sim 0.18$ at midrapidity, and shadowing effects are
small.  In this energy and rapidity regime, the $A$ dependence of $J/\psi$
production is attributed to nuclear absorption of the $J/\psi$: either the
pre-resonant $c \overline c$ pair or the $J/\psi$ itself can interact with
nucleons along its path and break up into charm hadrons in reactions
such as $p J/\psi
\rightarrow D \overline D X$ or $p J/\psi \rightarrow \overline D \Lambda_c$.
However, at energies such as those available at RHIC and the LHC, especially
away from midrapidity, much smaller $x$ values may be reached, making shadowing
more important relative to absorption.  

The $c \overline c$ pair that produces the $J/\psi$ can be created in
either color singlet or color octet states.
In Ref.~\cite{KSghfit}, absorption was described in terms of the singlet and
octet components of the $J/\psi$ wavefunction,
\begin{eqnarray}
|J/\psi \rangle = a_0 |(c \overline c)_1 \rangle + a_1 |(c \overline c)_8 g
\rangle + a_2 | (c \overline c)_1 gg \rangle + a_2' | (c \overline c)_8 gg
\rangle + \cdots \, \, .
\label{fockexp}
\end{eqnarray}
In the color singlet model \cite{baru}, 
only the first component is nonzero
for direct $J/\psi$ production.  The $c \overline c$ pairs pass
through nuclear matter in small color singlet states and reach their
bound state size outside the nucleus, except at sufficiently negative rapidity
where the $J/\psi$ may still hadronize inside the nuclear medium.
If $c \overline c$ pairs are predominantly produced in
color octet states, then the $|(c \overline c)_8 g \rangle$ state
interacts with nucleons.  The produced color octet $c \overline c$ can 
neutralize its color by a nonperturbative interaction with a gluon.
Since the $|(c \overline c)_8 g \rangle$ state is fragile, 
a gluon exchange between it and a nucleon
would separate the $(c \overline c)_8$ from the gluon
and break up the unbound octet \cite{KSghfit}.  If the $|(c
\overline c)_8 g \rangle$ state evolves without
interaction, such as in $pp$ collisions, the additional gluon is
absorbed by the octet $c \overline c$ pair, hence `evaporating' the color.
The color evaporation model (CEM) thence does not care about the 
relative coefficients in Eq.~(\ref{fockexp}).  On the other hand, the
NRQCD approach, applied to the total cross section in Ref.~\cite{benrot},
provides the leading coefficients in the expansion of the wavefunction
in Eq.~(\ref{fockexp}), encompassing both singlet and octet production
and absorption.  We calculate absorption of color singlets and color octets in
the CEM as well as the combination of the two in NRQCD.  The NRQCD approach
fixes the fraction of charmonium states produced in color singlets and 
color octets,
determining the rate of singlet and octet absorption in Eq.~(\ref{fockexp}).

The nuclear dependence of hard process production such as heavy quarks and
quarkonium in $AB$ collisions is typically parameterized as
\begin{eqnarray}
\sigma_{AB} = \sigma_{NN} (AB)^\alpha
\end{eqnarray}
where $A$ and $B$ can be either protons or nuclei and $\sigma_{NN}$ is the 
production cross section in a nucleon-nucleon collision.  At central values of 
rapidity, $y$, or Feynman $x$, $x_F$, where $x_F = 2p_z/\sqrt{S_{NN}} = 
2m_T \sinh y/\sqrt{S_{NN}}$, $m_T = \sqrt{p_T^2 + m^2}$ where $p_T$ is the
transverse momentum and $m$ is the particle mass,
the $A$ dependence of open charm production is, to a good
approximation, linear with $\alpha = 1$ \cite{e789}.  
However, for quarkonium production,
$\alpha < 1$ at midrapidity.  The most recent measurements by experiment
E866 at Fermilab \cite{e866} and NA50 at CERN \cite{NA50}
suggest $\alpha \sim 0.94-0.95$.  
Dependence on the kinematic variables $x_F$ and $p_T$ 
has been observed.  Typically $\alpha(x_F)$ decreases with increasing $x_F$
at $x_F > 0$.

It is as yet unknown if $\alpha$ is a strong function of energy, as predicted
by some absorption models \cite{GavVogt,hvqyr}.  If color singlet absorption
is at work, the absorption contribution should decrease with energy  because
the singlet state will stay small until far outside the target \cite{GavVogt}. 
On the other hand, if the absorption cross section depends on the $N J/\psi$
center of mass energy, at higher values of $\sqrt{S_{NN}}$ the average
$N J/\psi$ center of mass
energy increases, thus increasing the absorption cross section
\cite{hvqyr}. 
Since both initial and final-state
effects such as shadowing and absorption may be dependent 
on $\sqrt{S_{NN}}$,
empirically it would seem that $\alpha$ should be energy dependent.  At low
energies, most analyses have assumed that absorption by nucleons is the only 
contribution to the $A$ dependence.  While this may be a good approximation at
midrapidity, measured by NA50, away from the central region $\alpha$ is
$x_F$ dependent, as previously mentioned.  
Thus absorption alone is not enough to 
explain the $x_F$ dependence, as already noted a number of times, see 
{\it e.g.}\ Refs.~\cite{rv866,vbh1,hufsim}.  

Indeed, the characteristic shape
of $\alpha(x_F)$ at high $x_F$, $x_F \geq 0.25$, also cannot be explained
by shadowing alone \cite{rv866}.  However, at heavy ion colliders, the 
relationship between $x_F$, rapidity, and $\sqrt{S_{NN}}$ means that this
interesting $x_F$ region is pushed to very far forward rapidities.  At 
$\sqrt{S_{NN}} = 200$ GeV and $y = 2.5$, the forward edges of the PHENIX muon
arms at RHIC, $x_F \sim 0.19$ for a $J/\psi$ with $p_T = 0$ while at 
$\sqrt{S_{NN}} = 5.5$ TeV and $y = 4$, the forward edge of the ALICE muon arm
at the LHC, $x_F \sim 0.03$.  Therefore, at collider energies, a combination
of absorption and shadowing effects may be sufficient to address the
$J/\psi$ data.

In this paper, we discuss the combined effects
of shadowing and absorption both in minimum bias d$A$ collisions and
as a function of centrality at RHIC and the LHC.  We focus on d+Au collisions
at $\sqrt{S_{NN}} = 200$ GeV at RHIC and d+Pb collisions at $\sqrt{S_{NN}} = 
6.2$ TeV at the LHC.  While it is unclear whether $pA$ or
d$A$ collisions will be used at the LHC, one advantage of d$A$ is that the
energy is closer to that of Pb+Pb collisions at $\sqrt{S_{NN}} = 5.5$ TeV
whereas the $p$+Pb center of mass energy per nucleon would be 8.8 TeV.
Thus the d+Pb combination has been suggested as a baseline measurement at the
LHC \cite{aliceppr}. 

Our $J/\psi$ calculations generally employ the 
color evaporation model which treats all
charmonium production identically to $c \overline c$ production below
the $D \overline D$ threshold, neglecting color and spin \cite{HPC}.  
The leading order (LO) 
rapidity distributions of $J/\psi$'s produced in d$A$ collisions at impact 
parameter $b$ is
\begin{eqnarray} \frac{d\sigma}{dy d^2b d^2r}\!\!\! & = & \!\!\! 
2 F_{J/\psi} K_{\rm th} \int dz  dz'
\int_{2m_c}^{2m_D} \! M dM \left\{ F_g^d(x_1,Q^2,\vec{r},z) 
F_g^A(x_2,Q^2,\vec{b} - \vec{r},z')
\frac{\sigma_{gg}(Q^2)}{M^2} \right. \label{sigmajpsi} \\ & & 
\mbox{} \left. \!\!\!\!\!\!\!\!\!\! \!\!\!\!\!\!\!\!\!\! \!\!\!\!\!\!\!\!\!\! 
\!\!\!\!\!\!\!\!\!\! + \! \sum_{q=u,d,s} [F_q^d(x_1,Q^2,\vec{r},z) 
F_{\overline q}^A(x_2,Q^2,\vec{b} - \vec{r},z') +
F_{\overline q}^d(x_1,Q^2,\vec{r},z) F_q^A(x_2,Q^2,\vec{b} - \vec{r},z')] 
\frac{\sigma_{q \overline 
q}(Q^2)}{M^2} \right\}  \, \,  .
\nonumber
\end{eqnarray}
The partonic cross sections for LO gluon fusion and 
$q \overline q$ annihilation are given in
Ref.~\cite{combridge}, $M^2 = x_1x_2S_{NN}$ and $x_{1,2} =
(M/\sqrt{S_{NN}}) \exp(\pm y)$.  
The fraction of $c \overline c$ pairs below the $D
\overline D$ threshold that become $J/\psi$'s, $F_{J/\psi}$, is fixed
at next-to-leading order (NLO) \cite{HPC}.  Both this fraction and the
theoretical $K$ factor, $K_{\rm th}$, which scales 
the LO cross section to the NLO
value, drop out of the ratios.  The $K$ factor is independent 
of rapidity \cite{rvkfact}.  
While our calculations are LO, 
the shadowing ratios are independent of the order of the calculation
\cite{psidaprl}.  We use
$m_c=1.2$ GeV and $Q = M$ \cite{HPC} with the
MRST LO parton densities.  We will compare our minimum bias
CEM results with calculations employing the NRQCD approach and discuss any
differences in the resulting shadowing and absorption patterns. 

We assume that the nuclear parton densities, $F_j^A(x,Q^2,\vec r,z)$, 
are the product of
the nucleon density in the nucleus, $\rho_A(s)$, the nucleon parton density,
$f_j^N(x,Q^2)$, and a shadowing ratio, $S^j_{{\rm P},{\rm S}}
(A,x,Q^2,\vec{r},z)$,
where $\vec{r}$ and $z$ are the 
transverse and longitudinal location of the parton in position space. 
The first subscript, P, refers to the choice of shadowing parameterization,
while the second,  S, refers to the spatial dependence.
Most available shadowing parameterizations, including the ones used here, 
ignore effects in deuterium.  However, we take the proton and neutron
numbers of both nuclei into account.  Thus,
\begin{eqnarray}
F_i^d(x,Q^2,\vec{r},z) & = & \rho_d(s) f_i^N(x,Q^2) \label{fadeut} \\
F_j^A(x,Q^2,\vec{b} - \vec{r},z') & = & \rho_A(s') 
S^j_{{\rm P},{\rm S}}(A,x,Q^2,\vec{b} - \vec{r},z') 
f_j^N(x,Q^2) \, \, ,  \label{fanuc} 
\end{eqnarray}
where $s=\sqrt{r^2+z^2}$ and $s' = \sqrt{|\vec{b} - \vec{r}|^2 + z'^2}$.  
With no nuclear modifications, $S^i_{{\rm P},{\rm
S}}(A,x,Q^2,\vec{r},z)\equiv 1$.  The nucleon densities of the heavy
nucleus are assumed to be Woods-Saxon distributions with $R_{\rm Au} =
6.38$ fm and $R_{\rm Pb} = 6.62$ fm \cite{Vvv}.  We use the H\'{u}lthen
wave function \cite{hulthen} to calculate the deuteron density
distribution.  The densities are normalized so that $\int d^2r dz
\rho_A(s) = A$. We employ the MRST LO parton densities \cite{mrstlo}
for the free nucleon.

We have chosen shadowing parameterizations developed by two groups
which cover extremes of gluon shadowing at low $x$.  The Eskola {\it
et al.} parameterization, EKS98, is based on the GRV LO \cite{GRV}
parton densities.  Valence quark shadowing is identical for $u$ and
$d$ quarks.  Likewise, the shadowing of $\overline u$, $\overline
d$ and $\overline s$ quarks are identical at $Q_0^2$. 
Shadowing of the heavier flavor sea quarks,
$\overline s$ and higher, is, however, calculated and evolved separately
at $Q^2 > Q_0^2$.  The shadowing
ratios for each parton type are evolved to LO for $1.5 < Q < 100$ GeV
and are valid for $x \geq 10^{-6}$ \cite{EKRS3,EKRparam}.
Interpolation in nuclear mass number allows results to be obtained for
any input $A$.  The parameterizations by Frankfurt, Guzey and
Strikman (FGSo, the original parameterization, along with FGSh and FGSl for
high and low gluon shadowing) combine Gribov theory with hard
diffraction \cite{FGS}.  They are based on the CTEQ5M
\cite{cteq5} parton
densities and evolve each parton species separately to NLO for $2 < Q
< 100$ GeV.  Although the $x$ range is $10^{-5} < x < 0.95$, the
sea quark and gluon ratios are unity for $x > 0.2$.  The EKS98 valence
quark shadowing ratios are used as input since Gribov theory does not
predict valence shadowing.  The FGSo parameterization is available for four
different values of $A$: 16, 40, 110 and 206 while FGSh and FGSl also include
$A = 197$.  We use $A = 206$ for Au with FGSo and $A=197$ for FGSh and FGSl.

\begin{figure}[p] 
\setlength{\epsfxsize=0.65\textwidth}
\setlength{\epsfysize=0.35\textheight}
\centerline{\epsffile{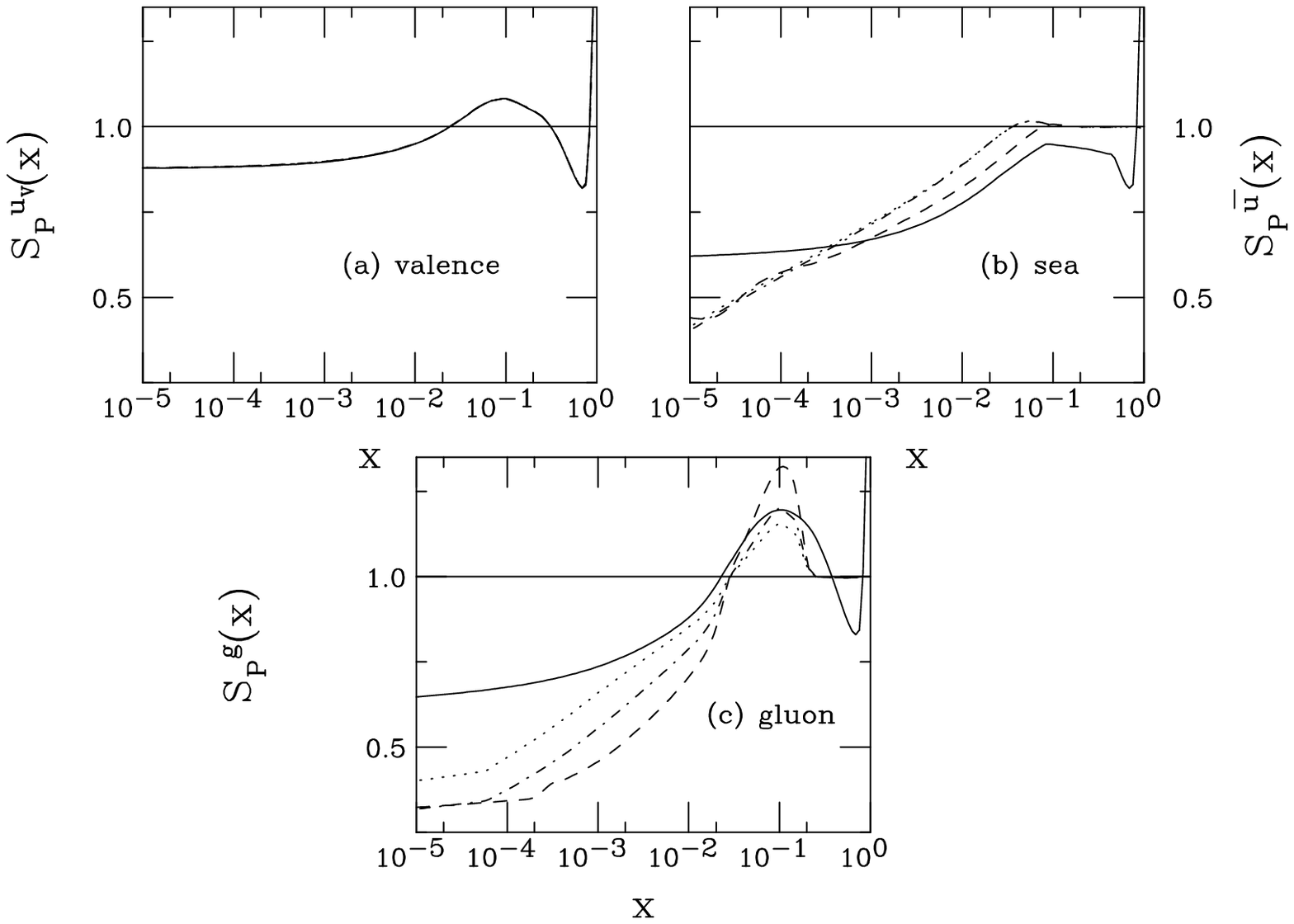}}
\caption[]{ The shadowing parameterizations are compared at
the scale $\mu = 2m_c = 2.4$ GeV.  The solid curves are EKS98, the dashed, 
FGSo, dot-dashed, FGSh, and dotted, FGSl.
}
\label{fshadow}
\end{figure}

Figure~\ref{fshadow} compares the four homogeneous ratios, $S^j_{\rm EKS}$
and $S^j_{\rm FGSi}$ for $Q=2m_c$.  The FGS calculation predicts more
shadowing at small $x$.  The
difference is especially large for gluons.  At very low $x$, the gluon
ratios for FGSo and FGSh are quite similar but in the intermediate $x$ regime,
the FGSh parameterization drops off more smoothly.  On the other hand, the
FGSl parameterization levels off at a higher value of $S_{\rm P}^i$ than the 
other two FGS parameterizations.  In the antishadowing 
regime, FGSh and FGSl are rather similar to EKS98 while FGSo has a larger 
antishadowing effect.  Obviously
shadowing alone will give an effective $A$ dependence as a function of 
rapidity with $y>0$ corresponding to low $x$, effectively mirroring the
curves in Fig.~\ref{fshadow}.

\begin{figure}[p]
\setlength{\epsfxsize=0.65\textwidth}
\setlength{\epsfysize=0.35\textheight}
\centerline{\epsffile{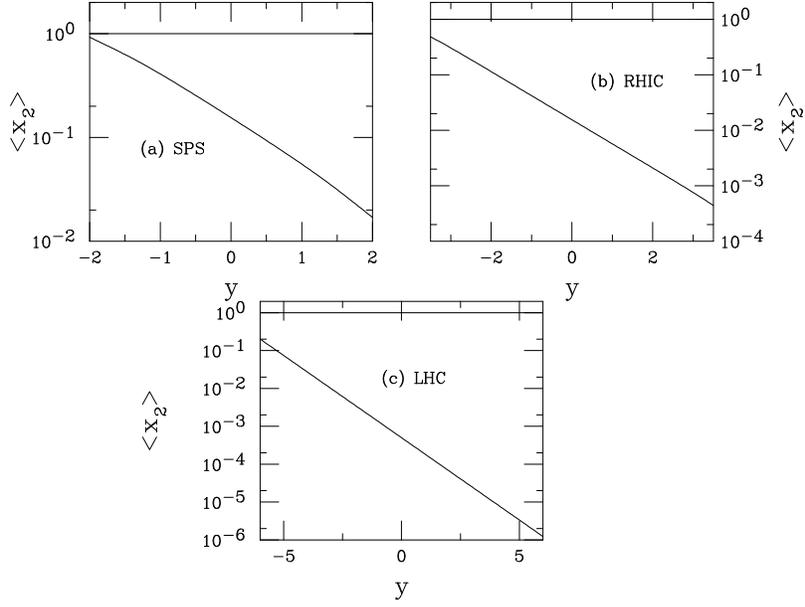}}
\caption[]{We give the average value of the nucleon momentum fraction, $x_2$,
of $J/\psi$ production in $pp$ collisions as a 
function of rapidity for (a) the CERN SPS with
$\sqrt{S} = 19.4$ GeV, (b) RHIC with $\sqrt{S} = 200$ GeV and (c) the LHC
with $\sqrt{S} = 6.2$ TeV.}
\label{xcomp}
\end{figure}

Figure~\ref{xcomp} shows the average value of $x_2$, the fraction of the
nucleon momentum carried by the interacting parton that produces the $J/\psi$,
for $pA$ collisions at the CERN SPS, $\sqrt{S_{NN}} = 19.4$ GeV, and d$A$
collisions at RHIC, $\sqrt{S_{NN}} = 200$ GeV, and the LHC, $\sqrt{S_{NN}} =
6.2$ TeV.  At midrapidity at the SPS, $\langle x_2 \rangle \sim 0.1$, in the
antishadowing region.  Note that if some level of antishadowing is indeed 
present at this energy, the effective absorption cross section determined from
minimum bias $pA$ data would actually be underestimated, as discussed in the
context of Pb+Pb collisions in Ref.~\cite{ekkv3}.
At midrapidity at RHIC, $\langle x_2 \rangle \sim 0.01$,
near the point where $S_{\rm P}^g \leq 1$.  In the forward region,
$\langle x_2 \rangle \sim 10^{-3}$ at $y \sim2$, 
clearly in the effective low $x$
regime due to $gg$ dominance while for negative rapidity, $y \sim -2$,
$\langle x_2 \rangle \sim 0.1$, in the antishadowing region.  Even lower values
of $\langle x_2 \rangle$ are reached in the measurable rapidity region of the
LHC, at $y \sim 4$, $\langle x_2 \rangle \sim 10^{-5}$, deep in the shadowing
region.  Systematic studies over the widest possible $\langle x_2 \rangle$
range can map out the gluon distribution, especially for the nucleus,
if absorption effects are understood.

We now turn to the spatial dependence of the shadowing, discussed in detail
in Ref.~\cite{psidaprl}.  Previously, when no inhomogeneous shadowing
parameterizations were available, we considered two forms of the spatial 
dependence, one proportional to the local nuclear density, $S^j_{{\rm P}, \,
\rm WS}$, and the other proportional to the parton path through the nucleus, 
$S^j_{{\rm P}, \, \rho}$.  Here we show only results for $S^j_{{\rm P}, \, 
\rho}$ and compare these
to the results with the inhomogeneous FGS parameterizations, FGSh and FGSl.
The $S^j_{{\rm P}, \, \rho}$ parameterization is, in general, \cite{psidaprl}
\begin{eqnarray}
S^j_{{\rm P},\, \rho}(A,x,Q^2,\vec r,z) = 1 + N_\rho 
(S^j_{\rm P}(A,x,Q^2) - 1)
\frac{\int dz \rho_A(\vec r,z)}{\int dz \rho_A(0,z)} 
\label{rhoparam} \, \, 
\end{eqnarray}
where $N_\rho$ is chosen so
that $(1/A) \int d^2r dz \rho_A(s) S^j_{{\rm P},\, \rho}(A,x,Q^2,\vec r,z) =
S^j_{\rm P}(A,x,Q^2)$. When $s = \sqrt{r^2 + z^2} \gg R_A$,
the nucleons behave as free particles while the modifications 
are larger than the average value $S^j_{\rm P}(A,x,Q^2)$ in
the center of the nucleus,.   
The normalization requires $(1/A) \int d^2r dz \rho_A(s) S^j_{{\rm
P},\, \rho} = S^j_{\rm P}$. 

While there are three homogeneous FGS parameterizations, no spatial 
dependence is provided for
FGSo, the case with the strongest gluon shadowing.  Therefore, we
also use $S^j_{{\rm P}, \, \rho}$ with this parameterization 
as well as with EKS98.  
The inhomogeneous FGSh and FGSl parameterizations are defined over $0.587\leq
s \leq 10$ fm for $A = 197$ and 208.  
Below $s_{\rm min} = 0.587$ fm, the ratios are fixed to
those at $s_{\rm min}$ while $S^j_{\rm FGSh,l}(A,x,Q^2,\vec s) 
\rightarrow 1$
for $s \geq 10$ fm. They do not consider the longitudinal spatial dimension.
The normalization requirement is similar to that of
Eq.~(\ref{rhoparam}), $(1/A) \int d^2s T_A(s) S^j_{\rm FGSh,l}(A,x,Q^2,\vec s) 
= S^j_{\rm FGSh,l}(A,x,Q^2)$.
Due to this fixed maximum, excluding the tail of the nuclear and deuteron
density distributions, these parameterizations will have a somewhat stronger
spatial dependence in a full d$A$ calculation.

\begin{figure}[p]
\setlength{\epsfxsize=0.5\textwidth}
\setlength{\epsfysize=0.3\textheight}
\centerline{\epsffile{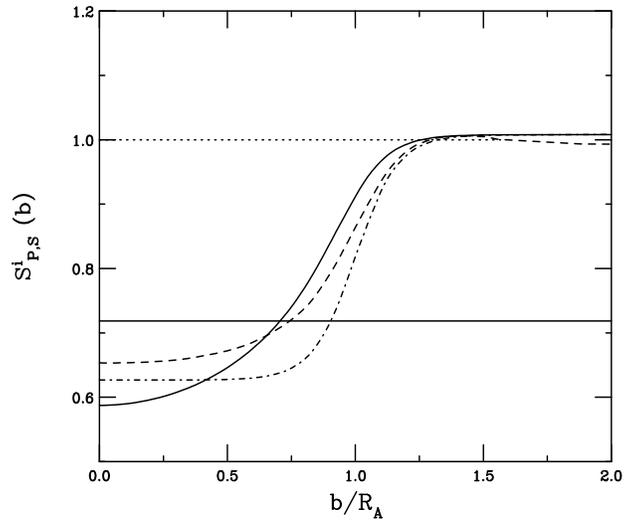}}
\caption[]{ The WS (dot-dashed) and $\rho$ (solid) inhomogeneous shadowing
parameterizations are compared to the inhomogeneous FGS shadowing
parameterization (dashed) at the same value of the homogeneous ratio, indicated
by the horizontal solid line.
}
\label{bshadcomp}
\end{figure}

Figure~\ref{bshadcomp} compares $S^j_{\rm FGSo,WS}$ and
$S^j_{{\rm FGSo},\rho}$ with $S^j_{\rm FGSh}(b)$ at
similar values of
the homogeneous shadowing ratios.  We see that $S^j_{{\rm FGSo},\rho}$ is quite
compatible with the available FGS inhomogeneous parameterizations.
 
\begin{figure}[p]
\setlength{\epsfxsize=0.95\textwidth}
\setlength{\epsfysize=0.3\textheight}
\centerline{\epsffile{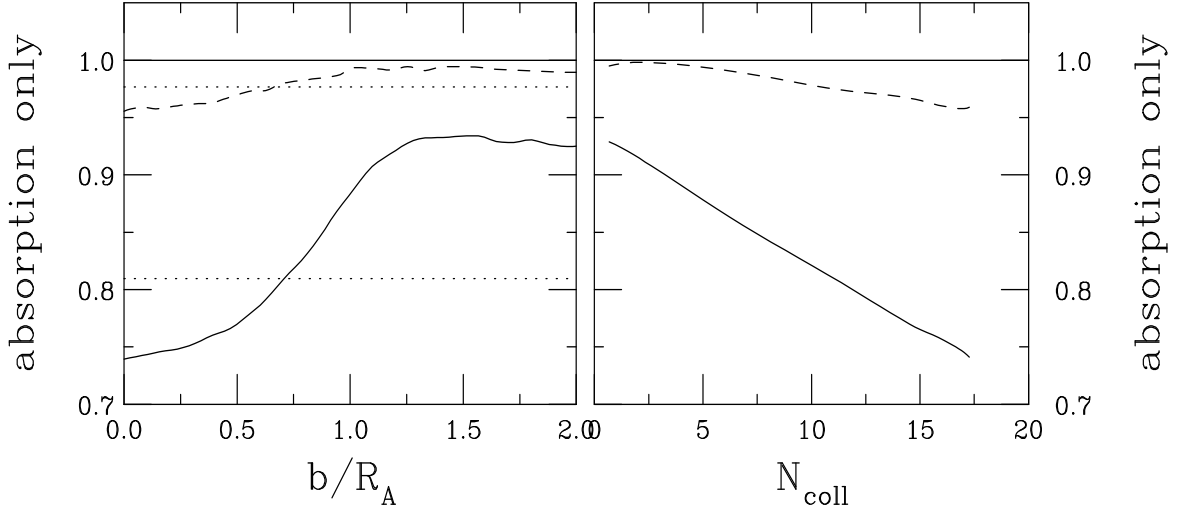}}
\caption[]{The $J/\psi$ dAu/$pp$ ratio for absorption alone with
$\sigma_{\rm abs} = 3$ mb as a function of impact parameter (left-hand side)
and as a function of the number of nucleon-nucleon collisions, $N_{\rm coll}$,
(right-hand side) for a constant octet (all $y$), solid, and singlet
($y = -2$), dashed.  The homogeneous results are indicated by the dotted lines.
}
\label{nucabs}
\end{figure}

To implement nuclear absorption on $J/\psi$
production in
d$A$ collisions, the production cross section in Eq.~(\ref{sigmajpsi})
is weighted by the
survival probability, $S^{\rm abs}$, so that 
\begin{eqnarray} 
S^{\rm abs}(\vec b - \vec s,z^\prime) = \exp \left\{
-\int_{z^\prime}^{\infty} dz^{\prime \prime} 
\rho_A (\vec b - \vec s,z^{\prime \prime}) 
\sigma_{\rm abs}(z^{\prime \prime} - z^\prime)\right\} \, \, 
\label{nsurv} 
\end{eqnarray}
where $z^\prime$ is the longitudinal production point, 
as in Eq.~(\ref{fanuc}), and 
$z^{\prime \prime}$ is the point at which the state is absorbed. 
The nucleon absorption cross section, $\sigma_{\rm abs}$, typically 
depends on where the
state is produced in the medium and how far it travels through nuclear matter.
If absorption alone is active, {\it i.e.}\ $S^j_{{\rm P}, \, {\rm S}} 
\equiv 1$, then an effective minimum bias $A$ dependence 
is obtained after integrating
Eqs.~(\ref{sigmajpsi}) and (\ref{nsurv}) over the spatial coordinates.
If $S^{\rm abs} = 1$ also, $\sigma_{{\rm d}A} = 2A \sigma_{pN}$.  
When $S^{\rm abs} \neq 1$, $\sigma_{{\rm d}A} = 2A^\alpha \sigma_{pN}$ where,
if $\sigma_{\rm abs}$ is a constant, independent of the production mechanism
for a nucleus of $\rho_A = \rho_0 \theta(R_A - b)$,  
\begin{eqnarray}
\alpha = 1 - \frac{9\sigma_{\rm abs}}{16 \pi r_0^2} \label{alfdef}
\end{eqnarray} 
where $r_0 = 1.2$ fm. 
Although we assume absorption is only effective for the heavy 
nucleus, the spatial dependence of the deuteron wavefunction is included in
Eq.~(\ref{sigmajpsi}).  The impact parameter dependence of absorption alone
is shown in Fig.~\ref{nucabs} for $\sigma_{\rm abs} = 3$ mb for a constant
octet cross section, independent of rapidity, and color singlet absorption 
at $y = -2$ where the $J/\psi$ can still hadronize in the target at 
$\sqrt{S_{NN}} = 200$ GeV.  The minimum bias results are indicated by the
dotted lines.  Absorption is stronger at central impact parameters but dies
away gradually at large values of $b$ due to the long tails of the density
distributions, particularly of deuterium.

The observed $J/\psi$ yield includes feed down from $\chi_{cJ}$ and $\psi'$
decays, giving
\begin{eqnarray}
S_{J/\psi}^{\rm abs}(\vec b - \vec s,z') = 0.58 S_{J/\psi, \, {\rm dir}}^{\rm 
abs}(\vec b - \vec s,z')
+ 0.3 S_{\chi_{cJ}}^{\rm abs}(\vec b - \vec s,z') + 0.12 S_{\psi'}^{\rm 
abs}(\vec b - \vec s,z') \, \, . \label{psisurv}
\end{eqnarray}
As discussed previously, in color singlet production, 
the final state absorption cross section depends on the size of the
$c \overline c$ pair as it traverses the nucleus, allowing absorption to be
effective only while the cross section is growing toward its asymptotic size
inside the target.
On the other hand, if the $c \overline c$ is produced as a color octet,
hadronization will occur only after the pair has traversed the target
except at very backward rapidity.  We have considered a constant octet cross
section, as well as one that reverts to a color singlet at backward rapidities.
For singlets, $S_{J/\psi, \, {\rm dir}}^{\rm abs} \neq
S_{\chi_{cJ}}^{\rm abs} \neq S_{\psi'}^{\rm abs}$ since $\sigma_{\rm abs}^{\psi
^\prime} > \sigma_{\rm abs}^{\chi_c} > \sigma_{\rm abs}^{J/\psi}$
but, with octets,
one assumes that $S_{J/\psi, \, {\rm dir}}^{\rm abs} = S_{\chi_{cJ}}^{\rm abs}
= S_{\psi'}^{\rm abs}$ since the asymptotic absorption cross section is 
predicted to be equal for all charmonium states.
Finally, we have also considered a combination of octet and
singlet absorption in the context of NRQCD.  See Ref.~\cite{rvherab}
for a detailed description of all the absorption models discussed.  

\begin{figure}[p]
\setlength{\epsfxsize=0.95\textwidth}
\setlength{\epsfysize=0.3\textheight}
\centerline{\epsffile{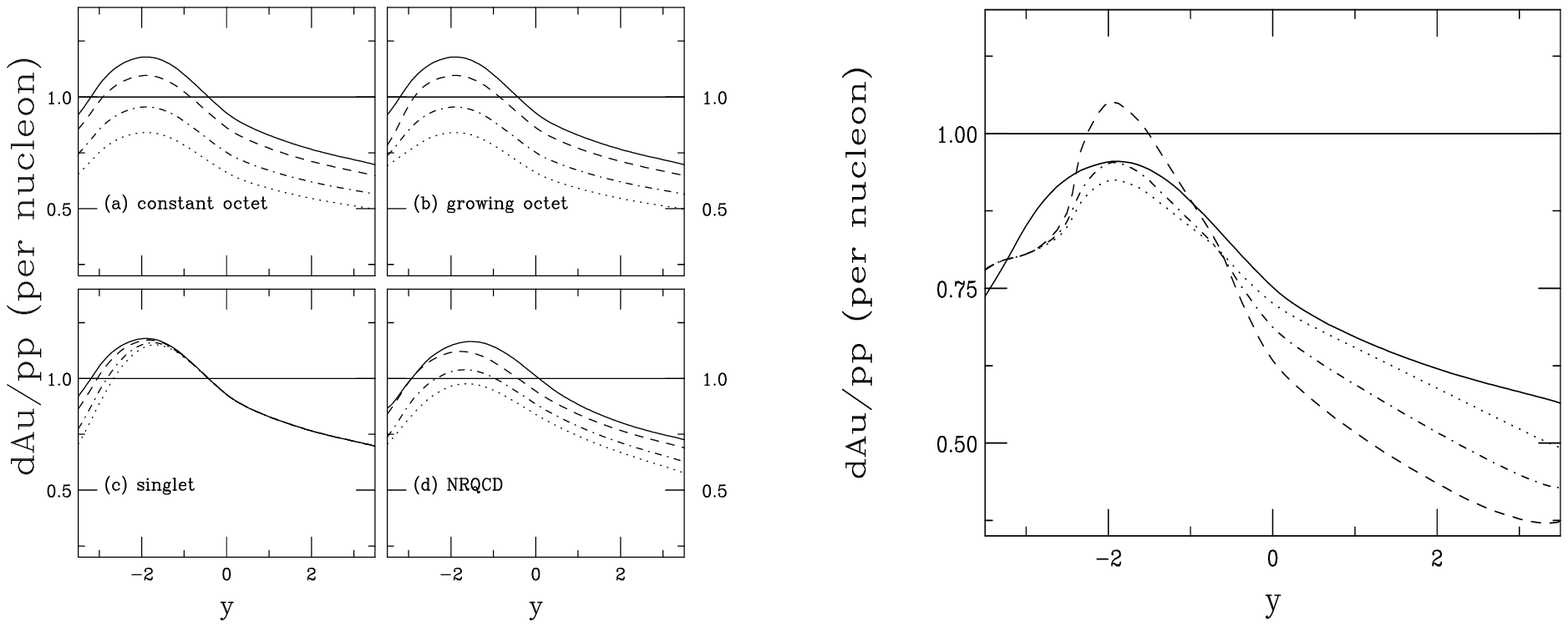}}
\caption[]{Left-hand side:  The $J/\psi$ dAu/$pp$ ratio with EKS98 at 200 GeV
as a function of rapidity for (a) constant octet, (b) growing
octet, (c)
singlet, all calculated in the CEM and (d) NRQCD.  For (a)-(c),
the curves are no
absorption (solid), $\sigma_{\rm abs} = 1$ (dashed), 3 (dot-dashed)
and 5 mb (dotted).  For (d), we show no absorption
(solid), 1 mb octet/1 mb singlet
(dashed), 3 mb octet/3 mb singlet (dot-dashed), and 5 mb octet/3 mb singlet
(dotted).  Right-hand side:  The $J/\psi$ dAu/$pp$ ratio at 200 GeV for a 
growing octet with $\sigma_{\rm abs} = 3$ mb is compared for four shadowing 
parameterizations.  We show the EKS98 (solid), FGSo (dashed), FGSh (dot-dashed)
and FGSl (dotted) results as a function of rapidity.
}
\label{abs_rhic}
\end{figure}

The left-hand side of Fig.~\ref{abs_rhic} shows EKS98 shadowing combined with
the absorption models described in the
text in minimum bias d+Au collisions at $\sqrt{S_{NN}} = 200$ GeV.
The difference between the constant and
growing octet cross sections
is quite small at large $\sqrt{S_{NN}}$ with only a minor octet-to-singlet 
conversion
effect at $y< -2$.  Color singlet absorption is also important only at similar
negative rapidities and is otherwise not different from shadowing alone.
The relative combination of nonperturbative singlet and octet contributions 
in NRQCD changes the shape of the shadowing ratio slightly.  The large 
$\chi_{cJ}$ singlet feeddown contribution reduces the overall absorption
effect on inclusive $J/\psi$ production.  Octet to singlet conversion, together
with singlet absorption, can be observed
at $y< -2$ as in the other cases.
 
Several values of the asymptotic
absorption cross section, $\sigma_{\rm abs} = 1$, 3 and
5 mb, corresponding to $\alpha = 0.98$, 0.95 and 0.92 respectively using
Eq.~(\ref{psisurv})
are shown in 
Fig.~\ref{abs_rhic}.  These values of
$\sigma_{\rm abs}$ are somewhat smaller than those obtained for the sharp
sphere approximation in Eq.~(\ref{alfdef}).  
The diffuse surface of a real nucleus and the longer range of the
density distribution results in a smaller value of $\sigma_{\rm abs}$ than
that found for a sharp sphere nucleus.  There is 
good agreement with the trend of the preliminary PHENIX data \cite{phenixqm04}
for $\sigma_{\rm abs} = 0-3$ mb.
We use a value of 3 mb in our further
calculations to illustrate the relative importance of absorption and shadowing.

The right-hand side of Fig.~\ref{abs_rhic} 
compares the EKS98 and FGS parameterizations, all with a growing octet
cross section with an asymptotic value of $\sigma_{\rm abs} = 3$ mb.
In the region that PHENIX can measure, the EKS98 and FGSl results are
essentially indistinguishable.  The FGSh result lies between the FGSo
and EKS98 results at forward rapidity but is also quite similar to EKS98
at negative rapidity.  

\begin{figure}[p]
\setlength{\epsfxsize=0.95\textwidth}
\setlength{\epsfysize=0.3\textheight}
\centerline{\epsffile{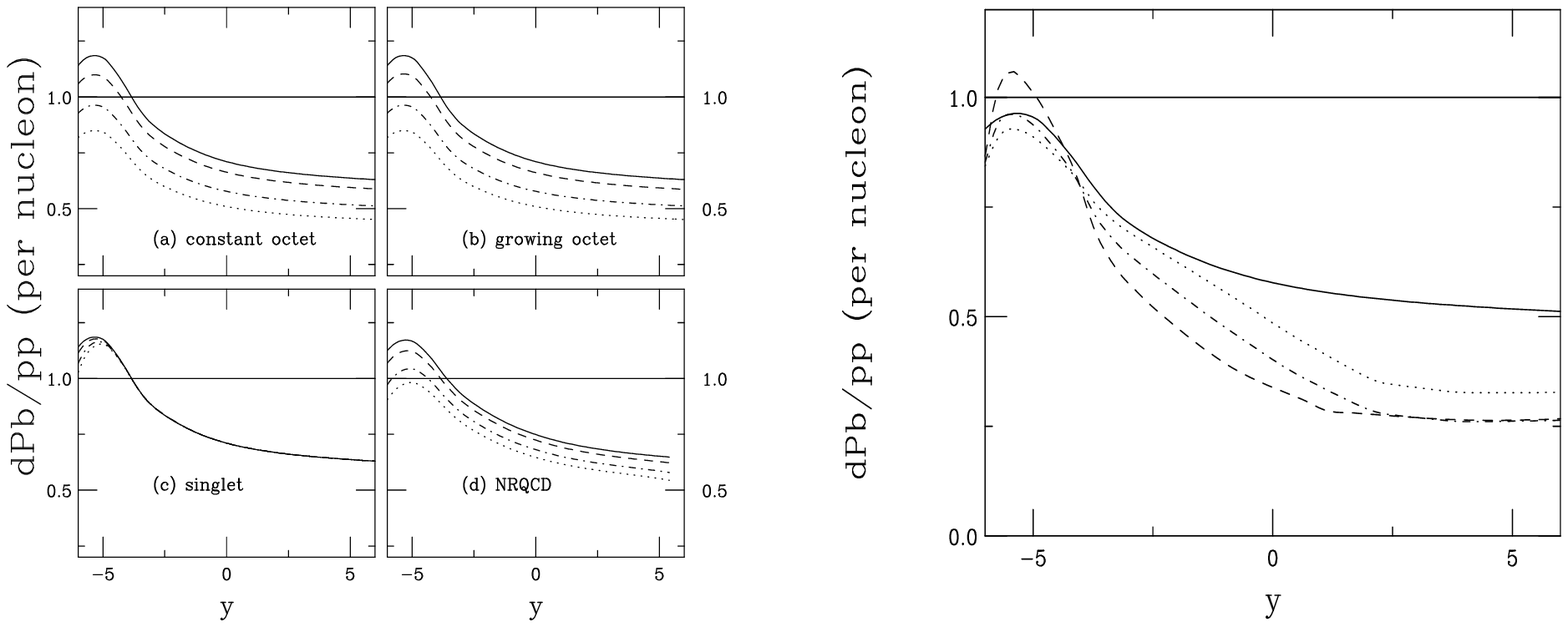}}
\caption[]{Left-hand side:  The $J/\psi$ dPb/$pp$ ratio with EKS98 at 6.2 TeV
as a function of rapidity for (a) constant octet, (b) growing
octet, (c)
singlet, all calculated in the CEM and (d) NRQCD.  For (a)-(c),
the curves are no
absorption (solid), $\sigma_{\rm abs} = 1$ (dashed), 3 (dot-dashed)
and 5 mb (dotted).  For (d), we show no absorption
(solid), 1 mb octet/1 mb singlet
(dashed), 3 mb octet/3 mb singlet (dot-dashed), and 5 mb octet/3 mb singlet
(dotted).  Right-hand side:  The $J/\psi$ dAu/$pp$ ratio at 200 GeV for a 
growing octet with $\sigma_{\rm abs} = 3$ mb is compared for four shadowing 
parameterizations.  We show the EKS98 (solid), FGSo (dashed), FGSh (dot-dashed)
and FGSl (dotted) results as a function of rapidity.
}
\label{abs_lhc}
\end{figure}

Figure~\ref{abs_lhc} shows the same calculations for d+Pb collisions at
$\sqrt{S_{NN}} = 6.2$ TeV at the LHC.  Although we have plotted the
results relative to $pp$ at the same energy, there are currently no
plans to run LHC $pp$ collisions at energies lower than 14 TeV.
Short, lower energy proton runs might better establish
the energy excitation functions for other processes so this possibility is
not excluded.  In any case, there
should be sufficient 14 TeV data to produce a high statistics $J/\psi$
baseline at this energy.  Between the Tevatron Run II data at 1.96 TeV
and the forthcoming LHC 
$pp$ data at 14 TeV, it should be possible extrapolate to
6.2 TeV with relative accuracy.  

In addition, it will not be possible to
directly compare these results to LHC data unless the entire $J/\psi$ $p_T$
spectrum can be measured since these calculations are 
integrated over $p_T$.  Tevatron Run II has shown that it is possible to 
measure the entire $J/\psi$ $p_T$ distribution at 
collider energies. 
A minimum $p_T$ cut would reduce the amount of shadowing due to the $Q^2$
dependence.

At 6.2 TeV, the difference between the constant and growing octet cross
sections is negligible for all rapidities shown while only a small absorption
effect is seen for the color singlet model at $y < -5$.  On the other hand,
the difference between
the shadowing parameterizations is larger.  Indeed, for $y>2$
there is nearly a factor of two difference between the EKS98 and FGS 
ratios, 0.5 and 0.25 respectively.  
This difference should be measurable in the ALICE muon arm, covering
the region $2.5<y<4$.  The results at negative rapidity are harder to 
discriminate since the difference between the shadowing ratios is decreased.
As the antishadowing region is approached, around $y = -5$, the difference
is almost negligible.

\begin{figure}[p]
\setlength{\epsfxsize=0.95\textwidth}
\setlength{\epsfysize=0.3\textheight}
\centerline{\epsffile{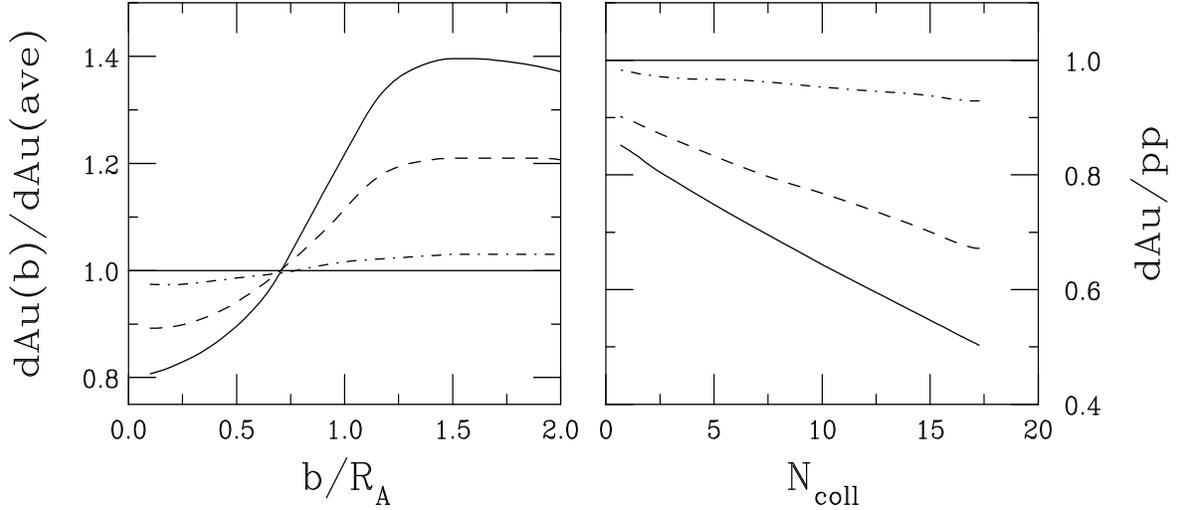}}
\caption[]{Left-hand side: The $J/\psi$ ratio dAu$(b)$/dAu(ave)
as a function of $b/R_A$.  Right-hand side:  The ratio dAu/$pp$ as a function
of $N_{\rm coll}$.  Results are shown for $y=-2$ (dot-dashed), $y=0$ (dashed) 
and $y=2$ (solid) at 200 GeV for a growing
octet with $\sigma_{\rm abs} = 3$ mb and the EKS98 parameterization.
}
\label{bdep_rhic}
\end{figure}

In central collisions, inhomogeneous shadowing is stronger than the 
homogeneous result.  The
stronger the homogeneous shadowing, the larger the inhomogeneity.  
In peripheral collisions, inhomogeneous
effects are weaker than the homogeneous results but some
shadowing is still present.  Shadowing persists in part because the
density in a heavy nucleus is large and approximately constant except
close to the surface and also because the deuteron wave function has
a long tail.  We also expect absorption to be a stronger effect in central
collisions.  In Fig.~\ref{bdep_rhic}, we show the inhomogeneous shadowing and
absorption results for EKS98 and $\sigma_{\rm abs} = 3$ mb at $\sqrt{S_{NN}} 
= 200$ GeV as a function of $b/R_A$ for the inhomogeneous to minimum bias
ratio 
(dAu$(b)$/$pp$)/(dAu(ave)/$pp) \equiv $dAu$(b)/$dAu(ave), shown on the 
left-hand side.  The ratios are shown 
for several values of rapidity to represent the
behavior in the anti-shadowing (large negative $y$), shadowing (large positive
$y$) and transition regions (midrapidity).  
The ratios are all less than unity for $b/R_A < 0.7$, 
with stronger than average shadowing and absorption, and rise above unity for 
large $b/R_A$, weaker than average shadowing and absorption. 

We note that for both color octet and color singlet production and absorption,
the spatial dependence of the absorption, shown in Fig.~\ref{nucabs}, is weaker
than the effect due to inhomogeneous shadowing at $x \sim 5\times
10^{-4}$ in Fig.~\ref{bshadcomp}.  The inhomogeneous shadowing 
effect has a stronger $y$ dependence than absorption which, at collider
energies, as illustrated in Figs.~\ref{abs_rhic} and \ref{abs_lhc}, is 
independent of $y$ except at large, negative rapidity.  This $y$-independence
can help disentangle the effects of shadowing and absorption as a function of
centrality.  For example, if the $b$ dependence of the dAu/$pp$ ratio at RHIC
could be determined at $y \sim -0.5$, where the result with $\sigma_{\rm abs} =
0$, shown in Fig.~\ref{abs_rhic}, indicates dAu/$pp \sim 1$, the effect of 
absorption alone could be fixed for all measurable $y$, allowing the effect
of shadowing alone to be determined for other values of $y$.

The right-hand side of Fig.~\ref{bdep_rhic}
shows the dAu/$pp$ ratios for the same rapidity values
as a function of the number of binary $NN$ collisions, $N_{\rm coll}$,
\begin{eqnarray}
N_{\rm coll}(b) = \sigma_{NN}^{\rm in} \int d^2s T_A(s)
T_B (|\vec b - \vec s|) \nonumber
\end{eqnarray}
where $T_A$ and $T_B$ are the nuclear thickness functions and $\sigma_{NN}^{\rm
in}$ is the inelastic nucleon-nucleon cross section, 42 mb at RHIC.  

The dependence of the ratios on $N_{\rm coll}$ is almost linear.  We do not
show results for $N_{\rm coll} < 1$, corresponding to $b/R_A > 1.3$ on the
left-hand side, the point where the dAu($b$)/dAu(ave) ratios begin to 
flatten out.  The weakest
$N_{\rm coll}$ dependence occurs in the antishadowing region, illustrated by
the $y = -2$ result (dot-dashed curve).  The trends of
the ratios as a function of $N_{\rm coll}$ are consistent with the PHENIX data
from the north muon arm ($y = 2$) and the electron arms ($y=0$)
but the preliminary
PHENIX results from the south arm ($y=-2$) are much stronger than
our predictions and, in fact, go the opposite way.  
The overall dependence on $N_{\rm coll}$ is stronger than that
obtained from shadowing alone, described in Ref.~\cite{psidaprl} where
inhomogeneous shadowing effects depend
strongly on the amount of homogeneous shadowing.  Relatively large effects
at low $x$ are accompanied by the strongest $b$ dependence.  In the
transition region around midrapidity at RHIC, the $b$ dependence of
the ratio dAu/$pp$ due to shadowing is nearly
negligible and almost all the $N_{\rm coll}$ dependence at $y \sim 0$
can be attributed to absorption.  The effect of absorption alone as a function 
of $N_{\rm coll}$ is shown on the right-hand side of Fig.~\ref{nucabs}.
The $y=-2$ results for color singlet production and absorption, 
in the antishadowing
region, are fairly independent of $N_{\rm coll}$.

\begin{figure}[p]
\setlength{\epsfxsize=0.95\textwidth}
\setlength{\epsfysize=0.5\textheight}
\centerline{\epsffile{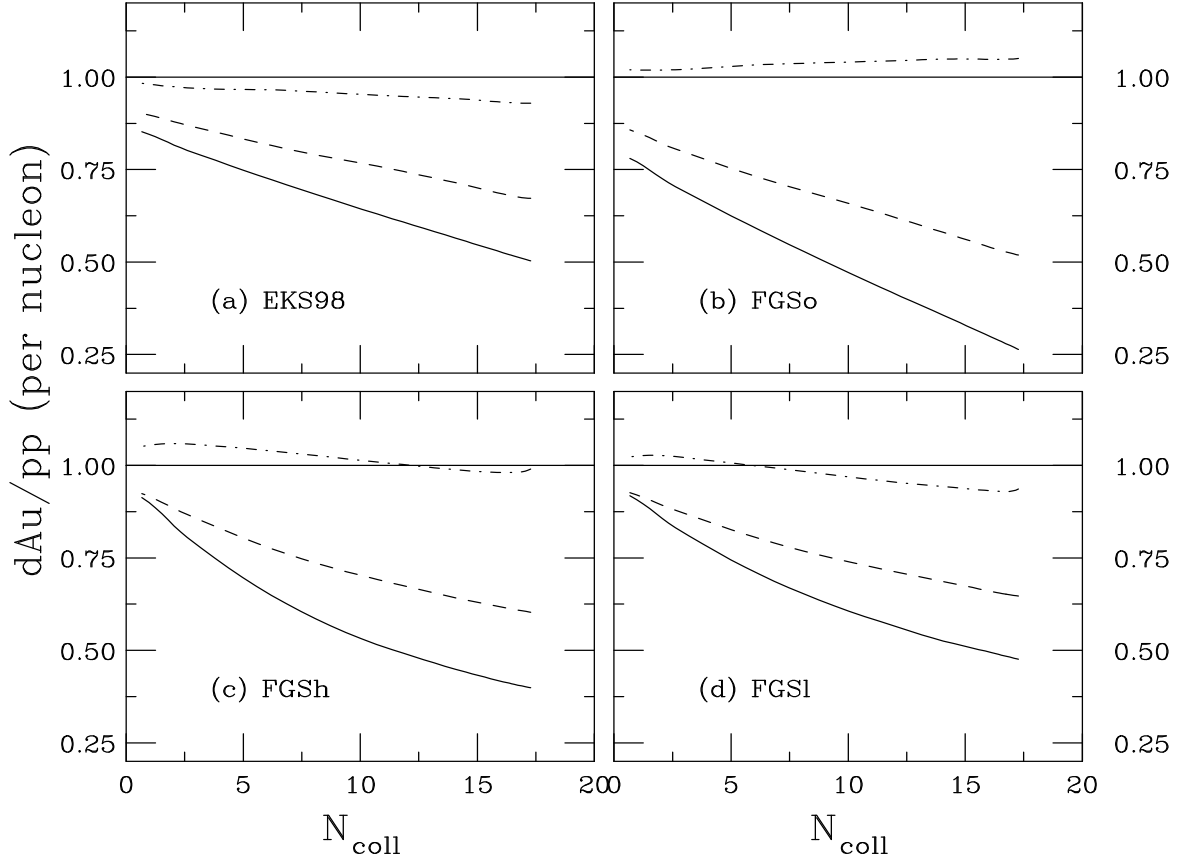}}
\caption[]{The ratio dAu/$pp$ as a function
of $N_{\rm coll}$ for the EKS98 (a), FGSo (b), FGSh (c) and FGSl (d) shadowing
parameterizations.  The calculations with EKS98 and FGSo use the inhomogeneous
path length parameterization while that obtained by FGS is used with FGSh and
FGSl.  Results are given for $y=-2$ (dot-dashed),
$y=0$ (dashed) and $y=2$ (solid) at 200 GeV for a growing
octet with $\sigma_{\rm abs} = 3$ mb.
}
\label{ncoll_rhic}
\end{figure}

Figure~\ref{ncoll_rhic} compares the EKS98 dAu/$pp$ ratios as a function 
of $N_{\rm coll}$ to results with the FGS parameterizations for the same
rapidities as in Fig.~\ref{bdep_rhic}.  Since EKS98 has the weakest
low $x$ shadowing, given the previous discussion
it is not surprising that it also has the weakest dependence
on $N_{\rm coll}$.  The FGSo results have the strongest $N_{\rm coll}$ 
dependence due to its strongest overall shadowing.  The FGSo ratio
goes above unity at $y =-2$ since, even with a 3 mb absorption cross section,
the minimum bias ratio is still greater than one at this rapidity.
While the FGSh and FGSl minimum bias ratios are intermediate to the EKS98
and FGSo ratios, they are more similar to EKS98 at forward rapidities so
that, especially for FGSl, the central results are quite similar to those
of EKS98 at all rapidities.  However, as $b$ increases and
$N_{\rm coll}$ decreases, they begin to differ.  Since the FGSh and FGSl
inhomogeneous parameterizations are tuned to become unity at 10 fm, the
shadowing component vanishes as $N_{\rm coll} \rightarrow 1$.  
On the other hand, there
is some residual absorption due to the overlapping tails of the deuteron and
nuclear density distributions. Thus the $y = 0$ and $y = 2$ curves meet at
$N_{\rm coll} = 1$.  At still higher impact parameters, where
$N_{\rm coll} < 1$, the
$y = -2$ curve intersects the higher rapidity curves.

While shadowing alone is not incompatible with the preliminary minimum
bias PHENIX data \cite{phenixqm04}, some
absorption should be present.  The 3 mb color octet
absorption cross section is also compatible with the minimum
bias data \cite{phenixsqm} and give reasonable agreement with the E866 data
when shadowing and other effects, important at forward $x_F$, are
included \cite{rv866}.  After the data is binned in centrality,
some discrimination between models may be possible.  Including inhomogeneous 
absorption steepens the dependence on $N_{\rm coll}$ more than 
with shadowing alone \cite{psidaprl}.  The preliminary PHENIX
centrality dependence \cite{phenixqm04} 
seems to be relatively flat for the central electron arms and the north 
muon arm (positive rapidity) while the centrality dependence in the south
muon arm (negative rapidity) seems to increase strongly with $N_{\rm coll}$.  
The EKS98 and FGSl calculations are most compatible with the central and north
muon arm data while FGSo and FGSh seem somewhat too strong relative to the
data.  Perhaps with a smaller absorption cross section, these calculations
could be made more compatible with the data.  None of the calculations 
agree with the preliminary
south muon arm data, exhibiting very little dependence on $N_{\rm coll}$.
The final PHENIX data as well as any data from a future d+Au run would
help better determine both the amount of gluon shadowing as well as the
relative contributions of shadowing and absorption, both in minimum bias
collisions and as a function of centrality.  It may be that the PHENIX data
would be more compatible with color singlet rather than color octet absorption 
although this would contradict fixed target results where shadowing is weak
\cite{NA50,e866}.

\begin{figure}[p]
\setlength{\epsfxsize=0.95\textwidth}
\setlength{\epsfysize=0.3\textheight}
\centerline{\epsffile{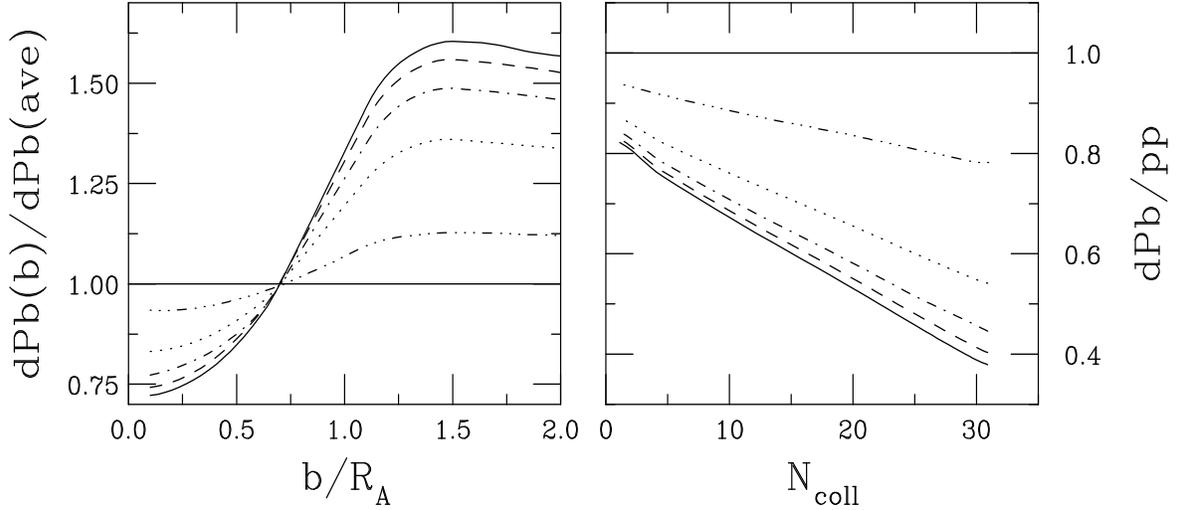}}
\caption[]{Left-hand side: The $J/\psi$ ratio dPb($b$)/dPb(ave)
as a function of $b/R_A$.  Right-hand side:  The ratio dPb/$pp$ as a function
of $N_{\rm coll}$.  Results are shown for $y = -4$ (dot-dot-dot-dashed),
$y=-2$ (dotted), $y=0$ (dot-dashed), $y = 2$ (dashed)
and $y=4$ (solid) at 6.2 TeV for a growing
octet with $\sigma_{\rm abs} = 3$ mb and the EKS98 parameterization.
}
\label{bdep_lhc}
\end{figure}

\begin{figure}[p]
\setlength{\epsfxsize=0.95\textwidth}
\setlength{\epsfysize=0.5\textheight}
\centerline{\epsffile{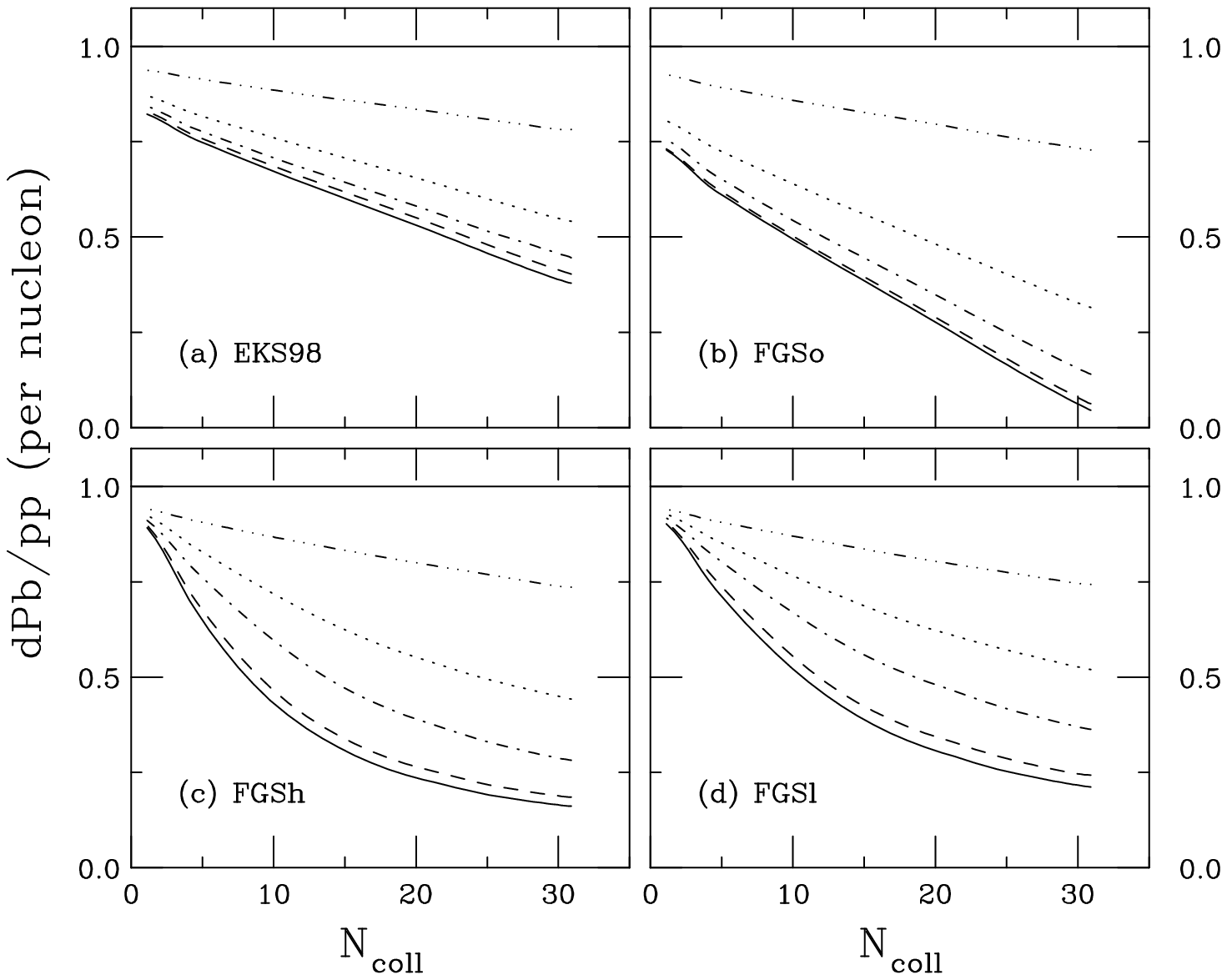}}
\caption[]{The ratio dPb/$pp$ as a function
of $N_{\rm coll}$ for the EKS98 (a), FGSo (b), FGSh (c) and FGSl (d) shadowing
parameterizations.  The calculations with EKS98 and FGSo use the inhomogeneous
path length parameterization while that obtained by FGS is used with FGSh and
FGSl.  Results are given for $y = -4$ (dot-dot-dot-dashed),
$y=-2$ (dotted), $y=0$ (dot-dashed), $y = 2$ (dashed)
and $y=4$ (solid) at 6.2 TeV  for a growing
octet with $\sigma_{\rm abs} = 3$ mb.
}
\label{ncoll_lhc}
\end{figure}

In Figs.~\ref{bdep_lhc} and \ref{ncoll_lhc} we present our inhomogeneous
shadowing and absorption calculations for d+Pb collisions at $\sqrt{S_{NN}} =
6.2$ TeV at the LHC.  Figure~\ref{bdep_lhc} shows only the EKS98 results
while Fig.~\ref{ncoll_lhc} compares the EKS98 results as a function of
$N_{\rm coll}$ to those of FGS.
Here we have also included results for $y = \pm 4$,
within the range of the ALICE muon arm.  Given that the rapidity range of
the muon arm encompasses the crossover point where dPb/$pp \sim 1$ at $y \sim
-3.9$, 
the centrality dependence of absorption alone could be determined and used to
calibrate the inhomogeneous shadowing effects.  Note that is is only
possible to reach $y \sim -3.9$ in ALICE by switching the beam directions and 
running Pb+d since the muon arm is only on one side of midrapidity.
Both ALICE and CMS should be able to
measure $J/\psi$ production at $y = \pm 2$ and 0.  
Since the exact $pp$ baseline is likely to be absent, as discussed above,
perhaps a better baseline for the centrality dependence would be the
ratio relative to the minimum bias result, shown on the left-hand side of
Fig.~\ref{bdep_lhc} as a function of $b/R_A$.
The change in dPb($b$)/dPb(ave) with
$b/R_A$ at the LHC is stronger than that of dAu($b$)/dAu(ave) at RHIC,
shown in Fig.~\ref{bdep_rhic}.

The results for $y = 4$, $\pm 2$ and 0, all in the low $x$ shadowing region,
are rather closely grouped together.  This should not be surprising because
the EKS98 shadowing ratios shown in Fig.~\ref{abs_lhc} are not very strong 
functions of rapidity.  The grouping of the results is even more striking
as a function of $N_{\rm coll}$, shown on the right-hand side of 
Fig.~\ref{bdep_lhc}.   Note also here that the $x$-axis scale is expanded
relative to that of RHIC since $\sigma_{NN}^{\rm in} = 76$ mb at 6.2 TeV,
increasing the number of inelastic $NN$ collisions possible at the LHC.
The expanded $N_{\rm coll}$ scale also has implications for the FGSh and FGSl
parameterizations with their sharp 10 fm cutoff on shadowing.  As seen in
Fig.~\ref{ncoll_lhc}(c) and (d), the curvature of these parameterizations is
much stronger than the linear dependence produced by Eq.~(\ref{rhoparam}).
This increased strength is due to the larger value of $\sigma_{NN}^{\rm in}$
at the LHC.  Now $N_{\rm coll} \approx 1$ at $b \sim 10$ fm.
Here FGSh and FGSl 
shadowing vanishes and the curves all come together at the value of the
ratio corresponding to the amount of residual absorption.  The curvature 
increases with $y$ since the larger shadowing ratio has to increase to unity
faster than those at negative rapidity where shadowing is weaker.

The results in Figs.~\ref{bdep_lhc} and \ref{ncoll_lhc} show that while,
at $y \leq 0$ in Fig.~\ref{abs_lhc} good statistical accuracy is necessary
to discriminate between shadowing models, the difference in their
$N_{\rm coll}$ dependence may be sufficient to do so.

\begin{figure}[p]
\setlength{\epsfxsize=0.95\textwidth}
\setlength{\epsfysize=0.3\textheight}
\centerline{\epsffile{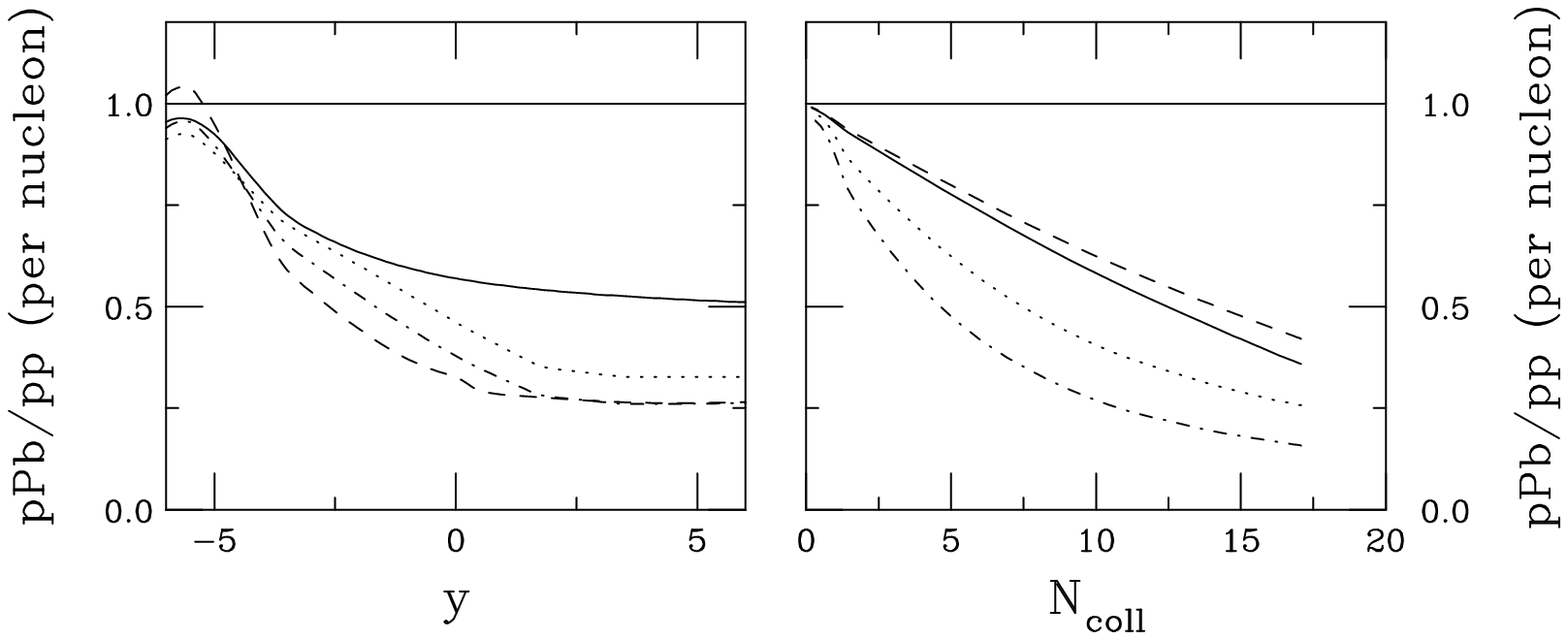}}
\caption[]{Left-hand side:  The ratio $p$Pb/$pp$ at 8.8 TeV as a function of 
rapidity for the EKS98 (solid), FGSo (dashed), FGSh (dot-dashed)
and FGSl (dotted) shadowing parameterizations with a 3 mb octet absorption
cross section.  Right-hand side:  The $N_{\rm coll}$ dependence for EKS98 ($y
= 4$, solid curve and $y = 0$, dashed curve) and FGSh ($y=4$, dot-dashed curve
and $y=0$, dotted curve).
}
\label{pa_lhc}
\end{figure}

Finally, since it is not yet clear whether proton or deuterium comparison
runs will be used at the LHC, Fig.~\ref{pa_lhc} shows the results for $p$+Pb
collisions at $\sqrt{S_{NN}} = 8.8$ TeV, the appropriate energy for 
proton-nucleus collisions.  We also have plotted the results assuming that
the center of mass rapidity is at $y=0$, as is the case for symmetric beams.
However, since the charge-to-mass, $Z/A$, ratio is different for protons and
lead, the proton beam has an energy of 7 TeV while the Pb beam has energy
2.75 TeV.  The asymmetric energies lead to a shift of the center of mass 
rapidity of 0.5 units in the direction of the proton beam.  
Since the difference between the $Z/A$ ratios of deuterium and lead is much
smaller, the shift in d+Pb collisions is 
only 0.1 unit.  The $p$+Pb antishadowing peak is shifted further 
toward negative rapidity while the region where the parameterizations flatten
begins closer to midrapidity than in Fig.~\ref{abs_lhc}.  Thus the EKS98
results as function of $N_{\rm coll}$, shown on the right-hand side of
Fig.~\ref{pa_lhc} for $y=4$ (solid curve) and $y=0$ (dashed curve) are quite
close together:  the shadowing ratio changes very little as a function of
rapidity.  The corresponding FGSh results are shown in the dot-dashed and
dotted curves respectively.  In both cases, the $N_{\rm coll}$ dependence is
stronger than in d+Pb collisions, as exhibited by the disappearance of both
shadowing and absorption as $N_{\rm coll} \rightarrow 1$ for the two 
inhomogeneous shadowing parameterizations.  The change in slope is particularly
strong for EKS98 because the proton is treated as a point particle 
while the deuteron is an extended system.  The maximum
$N_{\rm coll}$ for $p$+Pb is slightly larger than half that for d+Pb.  
If not for the
higher inelastic cross section, $\sigma_{NN}^{\rm in} = 81$ mb at 8.8 TeV, 
$N_{\rm coll}$ in $p$+Pb would be exactly half that of d+Pb.

In conclusion, the preliminary PHENIX data show that $J/\psi$ production in
d+Au collision is modified with respect to production in $pp$ collisions at the
same energy.  This modification is consistent with initial-state
shadowing plus final-state absorption seen in fixed-target experiments at 
800 GeV \cite{rv866}.  
More precise measurements may help to better set the level of absorption 
allowed by the data.  Precision measurements of the centrality dependence
may also help.
Corresponding data from the LHC will be more useful for separating and refining
shadowing models due the very low $x$ range available for large rapidity
measurements.  

We thank K.J. Eskola and V. Guzey for their shadowing parameterizations.
We also thank S.R. Klein and M.J. Leitch for discussions.
This work was supported in part by the Division
of Nuclear Physics of the Office of High Energy and Nuclear Physics of
the U.S. Department of Energy under Contract Number
DE-AC03-76SF00098.

\clearpage

\clearpage

\clearpage

\clearpage

\clearpage

\clearpage

\clearpage

\clearpage

\clearpage

\clearpage

\begin{references}

\bibitem{NA50} B. Alessandro {\it et al.} (NA50 Collaboration), Eur. Phys.
J. C {\bf 33}, 31 (2004).

\bibitem{Arn} J.J. Aubert {\it et al}., Nucl. Phys. B {\bf 293}, 740 (1987);
M. Arneodo, Phys. Rep. {\bf 240}, 301 (1994).

\bibitem{ekkv4}V. Emel'yanov, A. Khodinov, S.R. Klein and R. Vogt,
Phys. Rev. C {\bf 61}, 044904 (2000).

\bibitem{E745} T. Kitagaki {\it et al.}, Phys. Lett. B {\bf 214}, 281 (1988).

\bibitem{psidaprl} S.R. Klein and R. Vogt, Phys. Rev. Lett. {\bf 91}, 142301 
(2003).
 
\bibitem{KSghfit} D. Kharzeev and H. Satz, Phys. Lett. B {\bf 366}, 316 (1996).
 
\bibitem{baru} R. Baier and R. R\"{u}ckl, Z. Phys. C {\bf 19}, 251 (1983);
G.A. Schuler, hep-ph/9403387, CERN-TH.7170/94.
 
\bibitem{benrot} M. Beneke and I.Z. Rothstein, Phys. Rev. D {\bf 54}, 2005 
(1996).

\bibitem{e789} M.J. Leitch {\it et al.} (E789 Collaboration), Phys. Rev. Lett.
{\bf 72}, 2542 (1994).

\bibitem{e866} M.J. Leitch {\it et al.} (E866 Collaboration), Phys. Rev. Lett.
{\bf 84}, 3256 (2000).

\bibitem{GavVogt} S. Gavin and R. Vogt, Nucl. Phys. B {\bf 345}, 104 (1990).

\bibitem{hvqyr} M. Bedjidian {\it et al.}, hep-ph/0311048.

\bibitem{rv866}
R. Vogt,
Phys. Rev. C {\bf 61}, 035203 (2000).

\bibitem{vbh1}
R. Vogt, S.J. Brodsky, and P. Hoyer, Nucl. Phys. B {\bf 360}, 67 (1991).
  
\bibitem{hufsim}
J. H\"{u}fner and M. Simbel,
Phys. Lett. B {\bf 258}, 465 (1991).

\bibitem{aliceppr} P. Carminati {\it et al.} (ALICE Collaboration), J. Phys.
G {\bf 38}, 1517 (2004).

\bibitem{HPC} R.V. Gavai, D. Kharzeev, H. Satz, G. Schuler, K. Sridhar
and R. Vogt, Int. J. Mod. Phys. A {\bf 10}, 3043 (1995); G.A. Schuler
and R. Vogt, Phys. Lett. B {\bf 387}, 181 (1996).
 
\bibitem{combridge} B.L. Combridge, Nucl. Phys. B {\bf 151}, 429 (1979).

\bibitem{rvkfact} R. Vogt, Z. Phys. C {\bf 71}, 475 (1996); Heavy Ion Phys.
{\bf 17}, 75 (2003).

\bibitem{Vvv}
C.W. deJager, H. deVries, and C. deVries, Atomic Data and Nuclear Data 
Tables {\bf 14}, 485 (1974).

\bibitem{hulthen}D. Kharzeev, E.M. Levin and M. Nardi, hep-ph/0212316;
L. Hulthen and M. Sagawara, in {\it Handb\"uch der Physik}, {\bf 39}
(1957).

\bibitem{mrstlo} A.D.~Martin, R.G.~Roberts, and W.J. Stirling, and R.S. Thorne,
Phys. Lett. B {\bf 443}, 301 (1998).

\bibitem{GRV} 
M. Gl\"{u}ck, E. Reya, and A. Vogt, Z. Phys. C {\bf 53}, 127 (1992).

\bibitem{EKRS3}
K.J. Eskola, V.J. Kolhinen and P.V. Ruuskanen, Nucl. Phys. B {\bf 535}, 351 
(1998).

\bibitem{EKRparam}
K.J. Eskola, V.J. Kolhinen and C.A. Salgado, Eur. Phys. J. C {\bf 9}, 61 
(1999).

\bibitem{FGS} L. Frankfurt, V. Guzey and M. Strikman, arXiv:hep-ph/0303022.

\bibitem{cteq5} H.L. Lai {\it et al.}, Eur. Phys. J. C {\bf 12}, 375 (2000).

\bibitem{ekkv3}V. Emel'yanov, A. Khodinov, S.R. Klein and R. Vogt,
Phys. Rev. C {\bf 59}, 1860 (1999).

\bibitem{rvherab} R. Vogt, Nucl. Phys. A {\bf 700}, 539 (2002). 

\bibitem{phenixqm04} R. de Cassagnac (PHENIX Collaboration), J. Phys. G {\bf 
30}, S1341 (2004).
 
\bibitem{phenixsqm} H. Pereira (PHENIX Collaboration), proceedings of 
Strangeness in Quark Matter, Sept. 2004.

\end{references}
\end{document}